# Electronic transport in graphene-based heterostructures

J. Y. Tan[1,2], A. Avsar[1,2], J. Balakrishnan[1,2], G. K.W. Koon[1,2,3], T. Taychatanapat[1,2], E. C. T. O'Farrell[1,2], K. Watanabe[4], T. Taniguchi[4], G. Eda[1,2], A. H. Castro Neto[1,2], and B. Özyilmaz[1,2,3,a]

[1]Graphene Research Center, National University of Singapore, 117542, Singapore, [2]Department of Physics, National University of Singapore, 117542, Singapore, [3]NanoCore, National University of Singapore, 117576, Singapore, [4]National Institute for Materials Science, 1-1 Namiki, Tsukuba 305-0044, Japan

While boron nitride (BN) substrates have been utilized to achieve high electronic mobilities in graphene field effect transistors, it is unclear how other layered two dimensional (2D) crystals influence the electronic performance of graphene. In this letter, we study the surface morphology of 2D BN, gallium selenide (GaSe), and transition metal dichalcogenides (tungsten disulfide ($WS_2$) and molybdenum disulfide ($MoS_2$)) crystals and their influence on graphene's electronic quality. Atomic force microscopy analysis show that these crystals have improved surface roughness (root mean square (rms) value of only ~ 0.1nm) compared to conventional $SiO_2$ substrate. While our results confirm that graphene devices exhibit very high electronic mobility ($\mu$) on BN substrates, graphene devices on $WS_2$ substrates (G/$WS_2$) are equally promising for high quality electronic transport ($\mu$~38,000 $cm^2$/Vs at RT), followed by G/$MoS_2$ ($\mu$~10,000 $cm^2$/Vs) and G/GaSe ($\mu$~2,200 $cm^2$/Vs). However, we observe a significant asymmetry in electron and hole conduction in G/$WS_2$ and G/$MoS_2$ heterostructures, most likely due to the presence of sulphur vacancies in the substrate crystals. GaSe crystals are observed to degrade over time even under ambient conditions, leading to a large hysteresis in graphene transport making it a less suitable substrate.

[a]Author to whom correspondence should be addressed. Electronic mail:barbaros@nus.edu.sg

Isolated new 2D materials beyond graphene such as $WS_2$, $MoS_2$, and GaSe exhibit many exotic electronic[1], optical[2-4], spintronics[5] and mechanical[6] properties. Recent development in the transfer of these ultra thin planar structures onto each other with precise control offers outstanding opportunities for fundamental and applied studies. For example, the replacement of common silicon dioxide ($SiO_2$) substrate with ultra flat BN crystal resulted in a significant enhancement in electronic mobility of graphene that allowed the observation of room temperature (RT) ballistic transport[7-8] and fractional quantum Hall effect[9]. Combination of graphene with BN, $WS_2$ and $MoS_2$ crystals also enabled new memory and transistor concepts[10-11]. Recently, graphene and $WS_2$-based heterostructures have been demonstrated in highly flexible photovoltaic devices with extremely high external quantum efficiency[12]. Graphene based heterostructure devices built with $MoS_2$ or GaSe are also expected to exhibit similar behavior. Last but not least important, $WS_2$ was proposed to enhance the weak spin orbit coupling of graphene with a proximity effect[13]. Since graphene and these 2D crystals are the most active elements in current and possibly future heterostructure devices, it is important to understand the impact of these crystals on the electronic quality of graphene before building increasingly complex heterostructures.

In this letter, we study the electronic quality of graphene on various substrates. While $WS_2$ and $MoS_2$ have similar chemical and structural properties, G/$WS_2$ heterostructures exhibit four-fold higher electronic mobility than that of G/$MoS_2$, making $WS_2$ an attractive alternative to BN substrates. We observe conductivity saturation on electron side above a threshold voltage in G/$WS_2$ devices. GaSe crystals are found to be less inert to ambient conditions and graphene devices fabricated on GaSe crystals exhibit even lower mobilities than the graphene devices on

SiO$_2$ substrates. Our results demonstrate the importance of ideal choice of material for graphene-based heterostructure devices.

To study the surface morphology of crystals, micromechanical exfoliation method is employed to deposit relatively thick crystals of BN[14], WS$_2$[15], MoS$_2$ (Structure Probe Inc.-SPI, natural molybenite) and GaSe (HQ Graphene) on Si/SiO$_2$ wafers. Dark field imaging technique is implemented to select the potential candidate flakes of a ~300 μm$^2$ clean surface area and a height of ~20nm. Fig.1. (a-c) show the typical AFM images of BN, WS$_2$, and MoS$_2$ flakes after annealing in Ar/H$_2$ (9/1) gaseous mixture at 400°C for 6 hours to remove possible tape residues. We noticed that unlike other crystals, GaSe crystals corrode even under ambient conditions and corrode faster if the flakes are annealed or kept under strong intensity of light. The AFM images of GaSe flakes immediately after exfoliation and 1 day after exfoliation without annealing are shown in Fig. 1(d) and (e) respectively. The height histograms of the crystals are shown in Fig.1. (g) and corresponding rms roughness of corresponding crystals are summarized in Fig.1.(h). Fig.1. (f) is the AFM image of a conventional SiO$_2$ substrate for comparison purpose. BN has the flattest surface (~0.06nm) followed by WS$_2$ (~0.08nm), MoS$_2$ (~0.09nm), fresh GaSe (~0.12nm), and SiO$_2$ (~0.17nm). The roughness of a GaSe crystal increases from 0.12 nm to 0.185 nm after only one day even though the sample was kept in high vacuum, becoming rougher than SiO$_2$. Since the work functions of these 2D materials and graphene are similar, the charge neutrality point (CNP) of graphene is expected to be located at the center of band gap of these crystals, making them viable alternative substrate to SiO$_2$ for graphene field effect transistors[16].

The fabrication of graphene based heterostructure devices starts with micromechanical exfoliation of a graphene flake on a bilayer polymer stack adapted from Ref. 7. The bottom layer of the polymer stack is dissolved to isolate the remaining resist and the graphene from the supporting substrate. The resulting film is transferred onto previously exfoliated crystals. During this procedure, the graphene surface to be transferred onto the crystal surface never gets exposed to any solvent, resulting in ultra clean interfaces. Similar to previous reports on G/BN heterostructures[17-18], we observe bubble formation also at the interfaces of G/WS$_2$ and G/MoS$_2$ heterostructures. In order to minimize the effect of GaSe crystal degradation onto the electronic quality of graphene, graphene is transferred on GaSe immediately after it is characterized under optical microscope. Graphene is patterned with e-beam lithography into Hall bar structures in flat, bubble and wrinkle free regions and etched by O$_2$ plasma. The width and length of the graphene channels are 1 μm and 3 μm respectively. Finally Cr/Au (2 nm/100 nm) contacts are formed by thermal evaporation under high vacuum conditions. The devices are annealed at 340°C for 6 hours under Ar/H$_2$ gaseous mixture to minimize fabrication residues after etching and contact fabrication processes. Different from the rest, graphene on GaSe samples are fabricated without annealing. The optical image of a completed graphene device on WS$_2$ substrate is shown in Fig.1. (i). Transport measurements are performed with a four terminal ac lock-in technique under vacuum environment.

Compared to graphene on a SiO$_2$ substrate, graphene on a BN substrate has been already shown to have smaller impurity doping level and higher charge mobilities which is attributed to reduced surface roughness and surface charge traps[19-20]. We first performed charge characterization of a graphene device on BN substrate. Fig.2.(a) inset shows the back gate voltage (V$_{BG}$) dependence of the graphene resistivity ($\rho$) as determined from $\rho = \frac{Rw}{l}$ where $w$ is the width of graphene channel and $l$ is the spacing between electrodes. Charge neutrality point (CNP) is observed to be almost at V$_{BG}$ ~ 0, resistivity is below 50 Ω at the charge carrier density of $2 \times 10^{11}$ cm$^{-2}$ and full width at half maximum is extremely small (~0.5V) indicating that the sample has very high charge mobility. At the low density regime, a field effect mobility of ~300,000 cm$^2$/Vs (~190,000 cm$^2$/Vs) at 5K (300K) is extracted using $\mu = \frac{1}{e}\frac{d\sigma}{dn}$. This is consistent with the values previously reported.[7-8,19-20].

Next, we present the resistivity and conductivity of graphene on WS$_2$ substrate as a function of V$_{BG}$ at RT (Fig.2-a). Several features immediately distinguish the transport of this device from those on usual insulating substrates such as SiO$_2$. We observe that the conductivity is remarkably linear in V$_{BG}$ on hole side, however the sample exhibits a V$_{BG}$ independent conductivity above V$_{BG}$ ~ 45V on electron side. While reaching to the conduction channel of WS$_2$ could raise similar phenomena in graphene transport, the recent observation of hopping type transport in transition metal dichalcogenides due to the presence of high concentration of sulphur vacancies[21] suggests that these defect induced localized states act as a sink to the electronic charges of graphene once the Fermi level aligns with the level of localized states. For mobility discussion, we limit our analysis to the hole doped region. RT (6K) hole carrier mobilities of ~38,000 (46,000) cm$^2$/Vs at $5 \times 10^{11}$ cm$^{-2}$ and ~28,000 (30,500) cm$^2$/Vs at $3 \times 10^{12}$ cm$^{-2}$ are extracted. By using $\sigma^{-1} = ((ne\mu) + \sigma_0)^{-1} + \rho_s$, a density independent RT mobility of 35,000 cm$^2$/Vs is calculated. Electronic mobility of graphene on WS$_2$ is four times higher than on SiO$_2$ substrate and this makes WS$_2$ an appealing substrate for graphene to reach high mobilities. Even though we consistently get high mobilities, the position of the CNP is sample dependent. For the present sample, the CNP is located at V$_{BG}$ ~ 14V.

As a next step, we characterize the electronic quality of graphene on a MoS$_2$ substrate at RT. In all measured MoS$_2$ devices, we observe very high asymmetry in between electron and hole conductivity at RT (Fig.2.(b)). However, the electron conductivity of graphene on $MoS_2$ is not saturated within the $V_{BG}$ range applied. For the present device, a hole mobility of ~ 10,000 $cm^2$/Vs and an electron mobility of ~ 1,100 $cm^2$/V.s are calculated away from CNP. The sheet resistance in hole conduction side is ~250Ω which is higher than the value obtained for BN and $WS_2$ based graphene devices, but still comparable to the one on $SiO_2$ substrate[9]. A weak electron doping, possibly resulting from the impurities in $MoS_2$, is observed in the shown device[22].

While the surface roughness for $MoS_2$ and $WS_2$ are comparable, graphene on $MoS_2$ substrate has lower mobilities when compared to $WS_2$. We attribute this discrepancy to the lower thermal stability of $MoS_2$ leading to charge scattering from the oxidized $MoS_2$ surface. $MoS_2$ oxidation is reported to occur below 100 C in ambient conditions[23]. As mentioned previously, graphene based heterostructures are annealed several times at 340 C under gaseous mixture to remove the fabrication residues. We believe this annealing process results in the oxidation of the $MoS_2$ substrate and limiting the electronic performance of graphene on $MoS_2$. However the $WS_2$ surface is less prone to such oxidation compared to $MoS_2$ thus thermal annealing is more effective for achieving high mobilities on $WS_2$ samples[24,25]. For G/$MoS_2$ devices, instead of cleaning by annealing, mechanical cleaning can be adopted to improve the electronic quality of graphene while preventing the oxidation of the surface[26].

Finally, we measured graphene resistivity as a function of $V_{BG}$ on GaSe at 5 K (Fig 3-(a)). A field effect mobility of ~ 2,200 $cm^2$/Vs is extracted for the shown sample. The CNP of graphene is observed to be highly doped ($V_{BG}$ = 32 V). Similar low quality transport was observed in ultra flat graphene device on mica substrate and associated to the presence of charge traps on substrate[27-28]. In order to check the effect of GaSe charge traps onto the charge transport properties of graphene, we record the resistance while the $V_{BG}$ is swept forward (negative to positive) and backward (positive to negative) scans (Fig 3 (b)). We observe a significant hysteresis in resistivity and the hysteresis increases as $V_{BG}$ range increases. For example the hysteresis at a 50 V range is only ~ 2 V and it increases to ~37 V as the range increases to 90 V. The density of charge traps can be calculated by, $n_{tr} = C_g \Delta V / e$, where $\Delta V$ is the shift of the CNP with forward and backward sweeps, $C_g$ is the effective capacitance, ~ 718.5 e(μm)$^{-2}$V$^{-1}$, and e is the charge[29]. We calculated a charge trap density of 1.45x10$^{11}$ cm$^{-2}$ for $V_{BG}$=50 V range and 2.6x10$^{12}$ cm$^{-2}$ for 90 V range. The observed hysteresis in our system still persists even when the sample 1-) was measured at liquid helium temperatures. 2-) had undergone in-situ vacuum treatment at 1.5x10$^{-6}$ Torr for 72 hours, 3-) was in-situ annealed at 100°C for 6 hours under high vacuum conditions and 4-) post annealed at 340°C with a Ar/H$_2$ gaseous mixture in furnace. These observations exclude the water as a source of hysteresis[29-30]. As confirmed by our AFM characterization and recently discussed in a review by A. K. Geim et al[13], 2D GaSe flakes are not stable at ambient conditions and possibly creating the observed high density charged traps. Figure 3. (c) shows the dark field images of GaSe crystal just after exfoliation and 20, 40 and 80 seconds after capturing the first image. This degradation in ambient conditions is responsible for the observed high concentration of charged traps and this limits the electronic quality of graphene.

In summary, we have showed that 2D crystals have flatter surfaces than conventional SiO$_2$ substrates. Similar to previous reported values, we obtain very high electronic mobilities in G/BN heterostructures, followed by WS$_2$, MoS$_2$, and GaSe. Even though WS$_2$ has similar surface morphology with MoS$_2$, we observe higher mobilities in graphene devices on WS$_2$ than MoS$_2$. The observation of high density of charge traps on GaSe surface results in low mobility graphene. Our results demonstrate that the transport properties of graphene strongly depend on the underlying substrate and address the importance of choosing appropriately the active 2D crystals for heterostructure devices.

Note added in proof: During the preparation of this manuscript, we became aware of related works on the electronic transport in graphene-based heterostructures.[31,32]

B. Ö. would like to acknowledge support by the National Research Foundation, Prime Minister's Office, Singapore under its Research Fellowship (RF Award No. NRF-RF2008-7), and the SMF-NUS Research Horizons Award 2009-Phase II. A.H.C.N. would like to acknowledge support by the National Research Foundation, Prime Minister's Office, Singapore under its Competitive Research Programme (CRP Award No. NRF-CRP6-2010-5). G.E acknowledges Singapore National Research Foundation for funding the research under NRF Research Fellowship (NRF-NRFF2011-02).

Figure Captions

Fig. 1. Typical AFM scanning images of (a) BN, (b) WS$_2$, (c) MoS$_2$, (d-e) GaSe immediately after exfoliation and 1 day after exfoliation and (f) SiO$_2$. Height scale of the AFM image is 0-3 nm and scanning dimension is 1μmx1μm. (g-h) Height histogram and rms analysis of the images are shown in panels (a-f) respectively. (i) A completed graphene Hall bar device on WS$_2$ substrate.

Fig. 2. (a) Resistivity and conductivity of graphene on WS$_2$ substrate as a function of V$_{BG}$ at RT. Inset- V$_{BG}$ dependence graphene resistivity on BN substrate at 5K. (b) Resistivity and conductivity of graphene on MoS$_2$ substrate at RT.

Fig. 3. (a) Resistivity and conductivity of graphene on GaSe substrate as a function of V$_{BG}$. (b) Resistivity of graphene as a function of different V$_{BG}$ ranges. Black and red arrows represent the sweep directions from negative to positive and positive to negative, respectively. (c) Dark field images of GaSe crystal just after exfoliation and as a function of time.

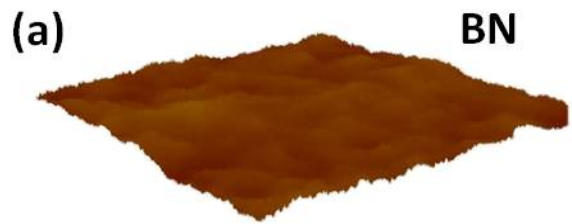 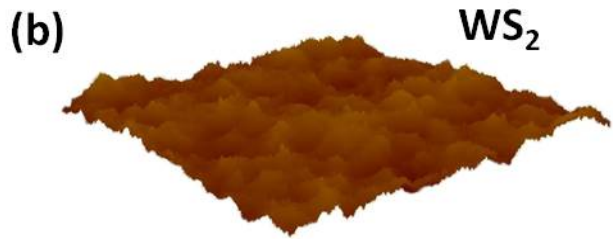 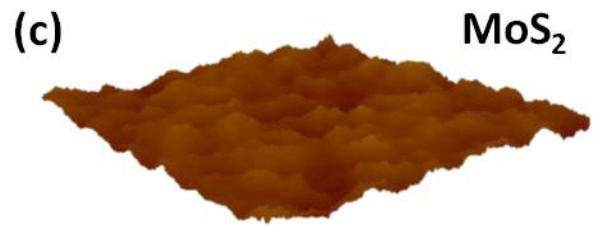
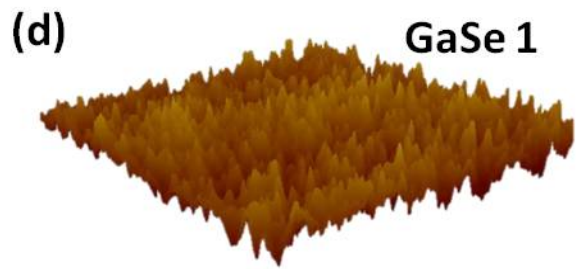 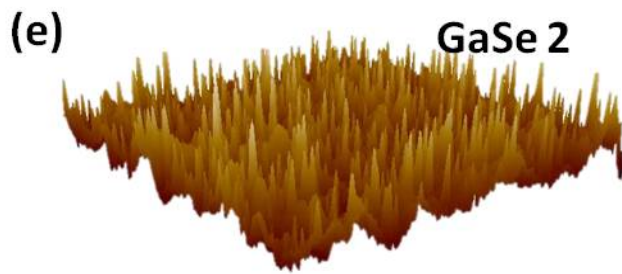 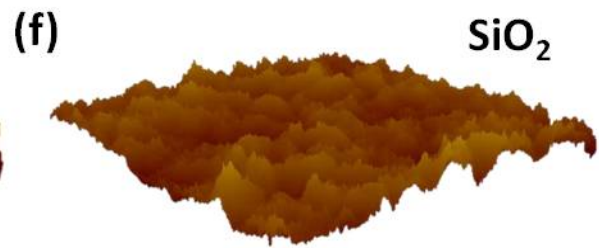
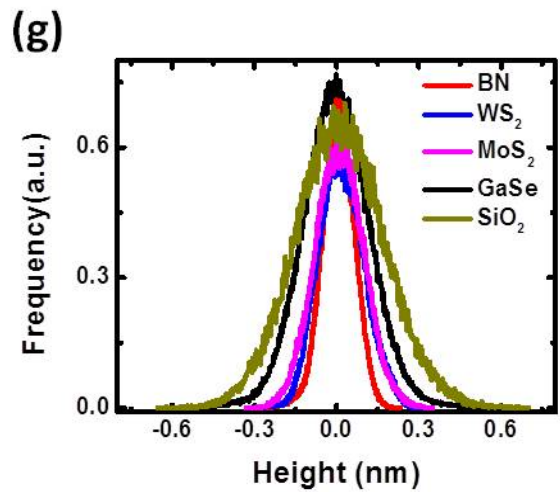 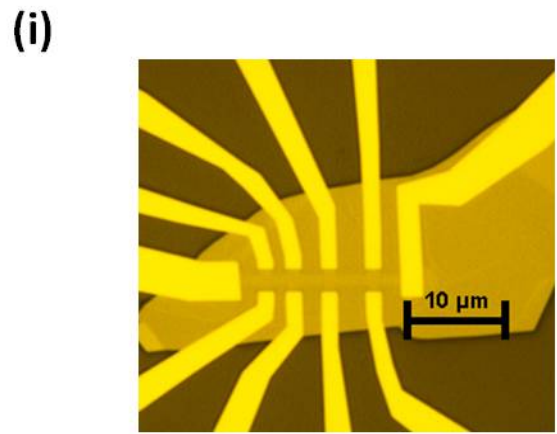

| Substrate | Roughness (nm) |
|---|---|
| BN | 0.06nm |
| WS$_2$ | 0.08nm |
| MoS$_2$ | 0.09nm |
| GaSe-1 | 0.12nm |
| GaSe-2 | 0.185nm |
| SiO$_2$ | 0.17nm |

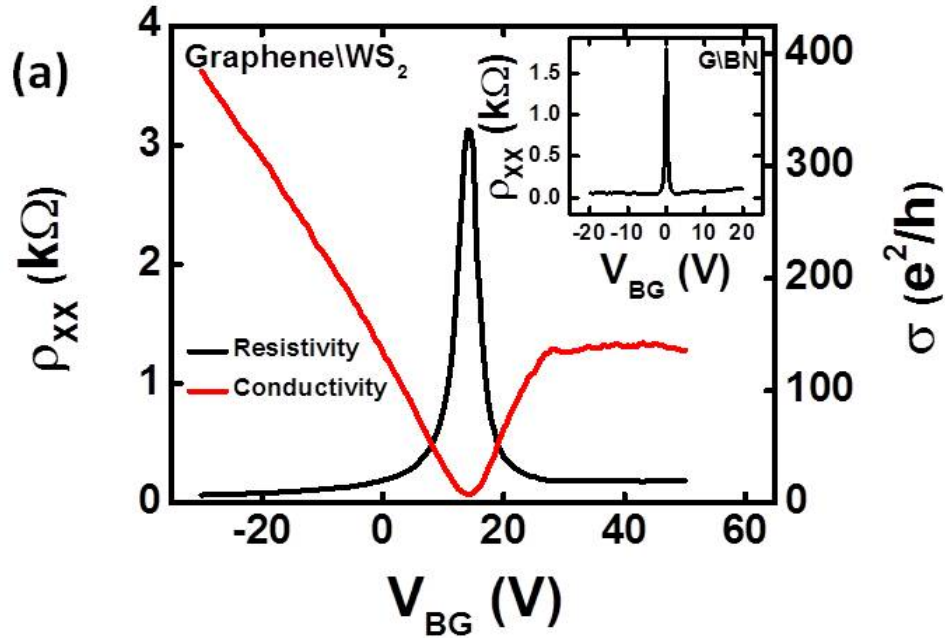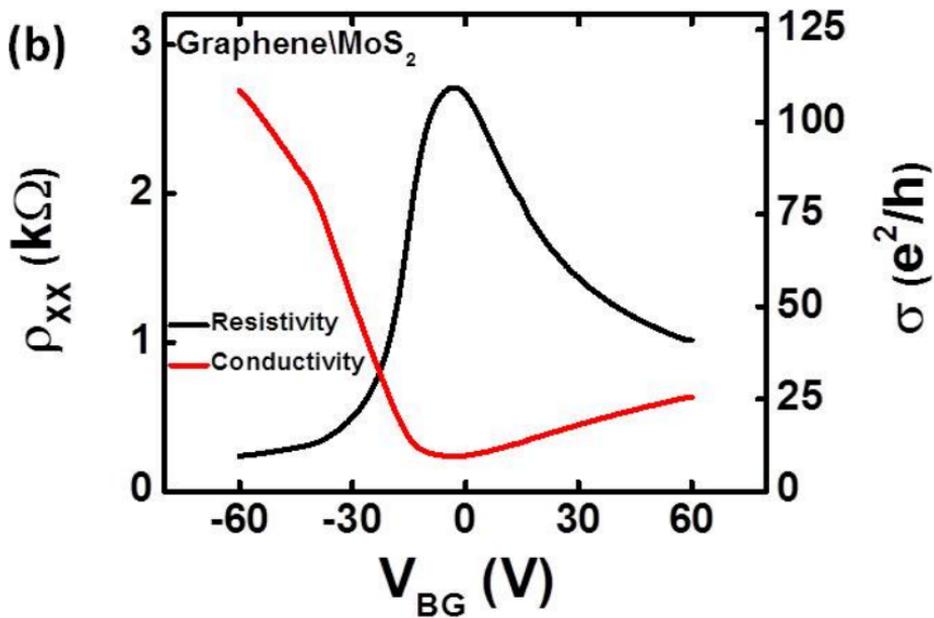

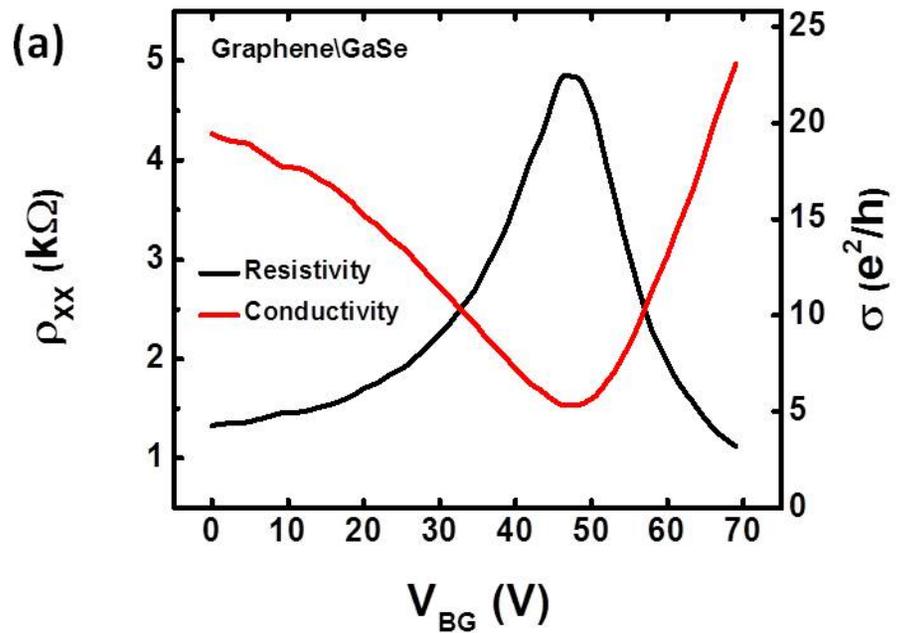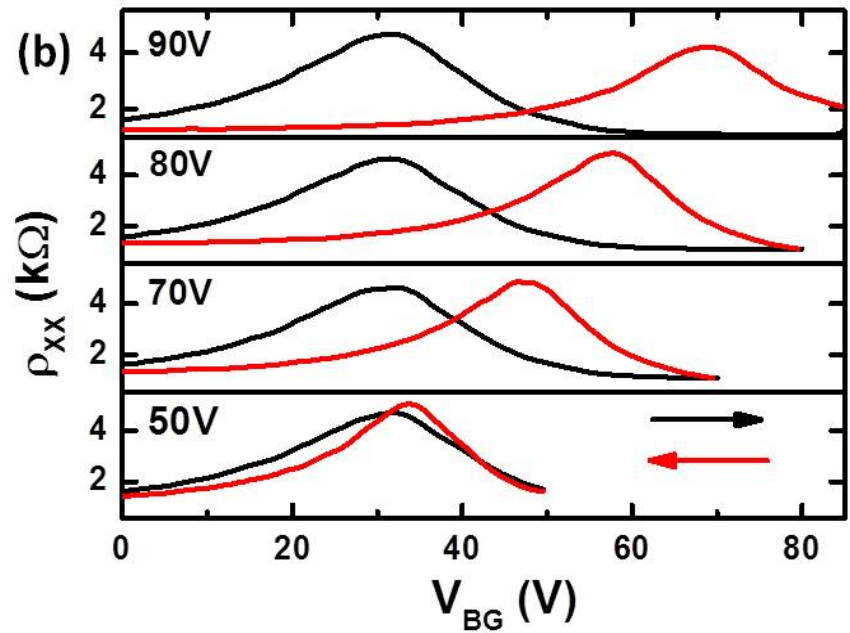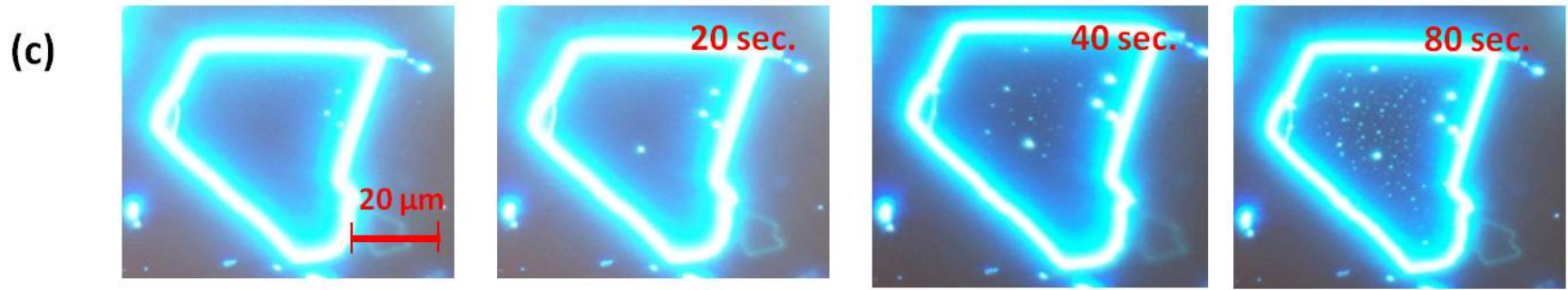